\documentclass[apj]{emulateapj}
\usepackage{apjfonts}

\shorttitle{New substellar companion to a massive young star}
\shortauthors{Lafreni\`{e}re et al.}

\newcommand{\msun}{\ensuremath{M_{\odot}}}
\newcommand{\mjup}{\ensuremath{M_{\rm Jup}}}
\newcommand{\teff}{\ensuremath{T_{\rm eff}}}
\newcommand{\primary}{HIP~78530}
\newcommand{\comp}{HIP~78530B}
\newcommand{\driftph}{{\sc Drift-Phoenix}}

\begin{document}

\title{Discovery of a $\sim$23~\mjup\ Brown Dwarf Orbiting $\sim$700 AU from the Massive Star HIP~78530 in Upper~Scorpius}

\author{David Lafreni\`ere\altaffilmark{1,2}, Ray Jayawardhana\altaffilmark{2}, Markus Janson\altaffilmark{2}, Christiane Helling\altaffilmark{3}, Soeren Witte\altaffilmark{4} \& Peter Hauschildt\altaffilmark{4}}
\altaffiltext{1}{D\'epartement de physique, Universit\'e de Montr\'eal, C.P. 6128 Succ. Centre-Ville, Montr\'eal, QC, H3C 3J7, Canada}
\altaffiltext{2}{Department of Astronomy and Astrophysics, University of Toronto, 50 St. George Street, Toronto, ON, M5S 3H4, Canada}
\altaffiltext{3}{SUPA, School of Physics and Astronomy, University of St. Andrews, North Haugh, St. Andrews KY16 9SS, UK}
\altaffiltext{4}{Hamburger Sternwarte, Gojenbergsweg 112, 21029 Hamburg, Germany}
\email{david@astro.umontreal.ca}

\begin{abstract}
We present the discovery of a substellar companion on a wide orbit around the $\sim$2.5~\msun\ star \primary, which is a member of the 5~Myr-old Upper Scorpius association. We have obtained follow-up imaging over two years and show that the companion and primary share common proper motion. We have also obtained $JHK$ spectroscopy of the companion and confirm its low surface gravity, in accordance with the young age of the system. A comparison with \driftph\ synthetic spectra indicates an effective temperature of $2800\pm200$~K and a comparison with template spectra of young and old dwarfs indicates a spectral type of M8$\pm$1. The mass of the companion is estimated to be 19-26~\mjup\ based on its bolometric luminosity and the predictions of evolutionary models. The angular separation of the companion is 4.5\arcsec, which at the distance of the primary star, 156.7~pc, corresponds to a projected separation of $\sim$710~AU. This companion features one of the lowest mass ratios ($\sim$0.009) of any known companion at separations greater than 100~AU. 
\end{abstract} 

\section{Introduction}

The recent direct imaging discoveries of very low mass substellar companions to stars -- especially those orbiting the young stars 1RXS J160929.1-210524 \citep{lafreniere08b, lafreniere10}, HR~8799 \citep{marois08}, Fomalhaut \citep{kalas08}, AB~Pic \citep{chauvin05b}, HN~Peg \citep{luhman07}, and CT~Cha \citep{schmidt08} -- have raised new questions about the formation mechanisms of planets and brown dwarfs. These companions have separations of tens to several hundred AU and mass ratios $\lesssim$0.02 relative to their primaries. Several theoretical ideas are under discussion to explain such properties, but they all run into some difficulties. Core accretion models favor formation of giant planets close to the 'snow line', i.e., at $<$10 AU \citep{pollack96}. Gravitational instability could produce massive planets at large radii, but would require unusually large and massive circumstellar disks \citep[e.g.][]{vorobyov10}. Fragmentation of pre-stellar cores during collapse does not lead to such extreme binary systems easily \citep[e.g.][]{bate03}. Thus, some theorists have recently proposed that dynamical instabilities within planetary systems that originally formed multiple giant planets could scatter some of them to large separations \citep{scharf09, veras09}. However, the more massive planets are less likely to reach the widest orbits. 

Here we report on the direct imaging discovery, common proper motion confirmation and multi-band spectroscopy of a $\sim$23~\mjup\ companion seen $\sim$700~AU from \primary, which is a $\sim$2.5~\msun\ star (spectral type B9V) in the Upper Scorpius young association. The separation of this new companion places it among the widest known substellar companions to stars and its extreme mass ratio -- among the lowest currently known -- is comparable to those of directly imaged planets, even though its mass is well above the deuterium-burning threshold.  Together with the companions mentioned above, this new substellar companion presents a good challenge to all formation scenarios and contributes to blurring the distinction between giant planets and brown dwarfs even further.

\section{Observations and data reduction}\label{sect:obs}

\subsection{Imaging}

The discovery presented in this paper was made as part of a direct imaging search for new stellar and substellar companions around about 90 stars in the Upper Scorpius region. 
The overall target sample was built by randomly selecting, from the list of Upper Scorpius stars in \citet{carpenter06}, an equal number of stars in each of five equal logarithmic mass bins over the range $\sim$0.15-5~\msun; the spectral types range from B0 to M5. The observations were made using the NIRI camera \citep{hodapp03} and the ALTAIR adaptive optics system \citep{herriot00} at the Gemini North Telescope. Except for a few faint targets, the target stars themselves were used for wave front sensing. The field lens of ALTAIR was used to reduce the effect of anisoplanatism. The first epoch imaging of \primary\ was done on 2008 May 24 in the narrow band filter $K_{\rm continuum}$ centered at 2.0975~$\mu$m. For sky subtraction we used 5 dither positions corresponding to the corner and center of a square of side 10\arcsec. At each position we obtained one co-addition of twelve 0.5~s integrations in fast, high read-noise mode, followed by one single 10~s integration in slow, low read-noise mode. At each position this provides an unsaturated image of the target star and a much deeper image of the field that can be readily spatially registered and scaled in flux. The full-width-at-half-maximum for this data set is 0.075\arcsec\ and the Strehl ratio is 0.34.

The initial image of \primary\ revealed an interesting faint nearby source which, based on a comparison with observations made by \citet{kouwenhoven05, kouwenhoven07}, was very likely to be a true co-moving companion. To confirm the common proper motion of this candidate and get additional photometry measurements, follow-up imaging observations in $J$, $H$ and $K^\prime$  were obtained on 2009 July 2 using the same instrument as before. To avoid saturation in these broad filters, we used a $512\times512$ subarray, and kept the strategy of a short, unsaturated image immediately followed by a longer, saturated image at each of 5 dither positions. For each wavelength, three sky frames were obtained at offsets of $>15\arcsec$. The short exposure times were 0.055~s with 40 co-additions, while the long exposures were 6~s, 5~s, and 4~s with one co-addition in $J$, $H$ and $K^\prime$, respectively. To increase the significance of our common proper motion confirmation further, another follow-up observation was made on 2010 August 30 using the  $K_{\rm cont}^{2.09}$ filter, full frame, and one co-addition of twelve 0.6~s integrations followed by a single 10~s integration at each of five dither positions. Other follow-up data were acquired in spring and summer 2010 but owing to an overseen difference in the instrumental setup, these data suffer from large systematic astrometric errors and are not used here.

\begin{figure*}
\epsscale{1}
\plotone{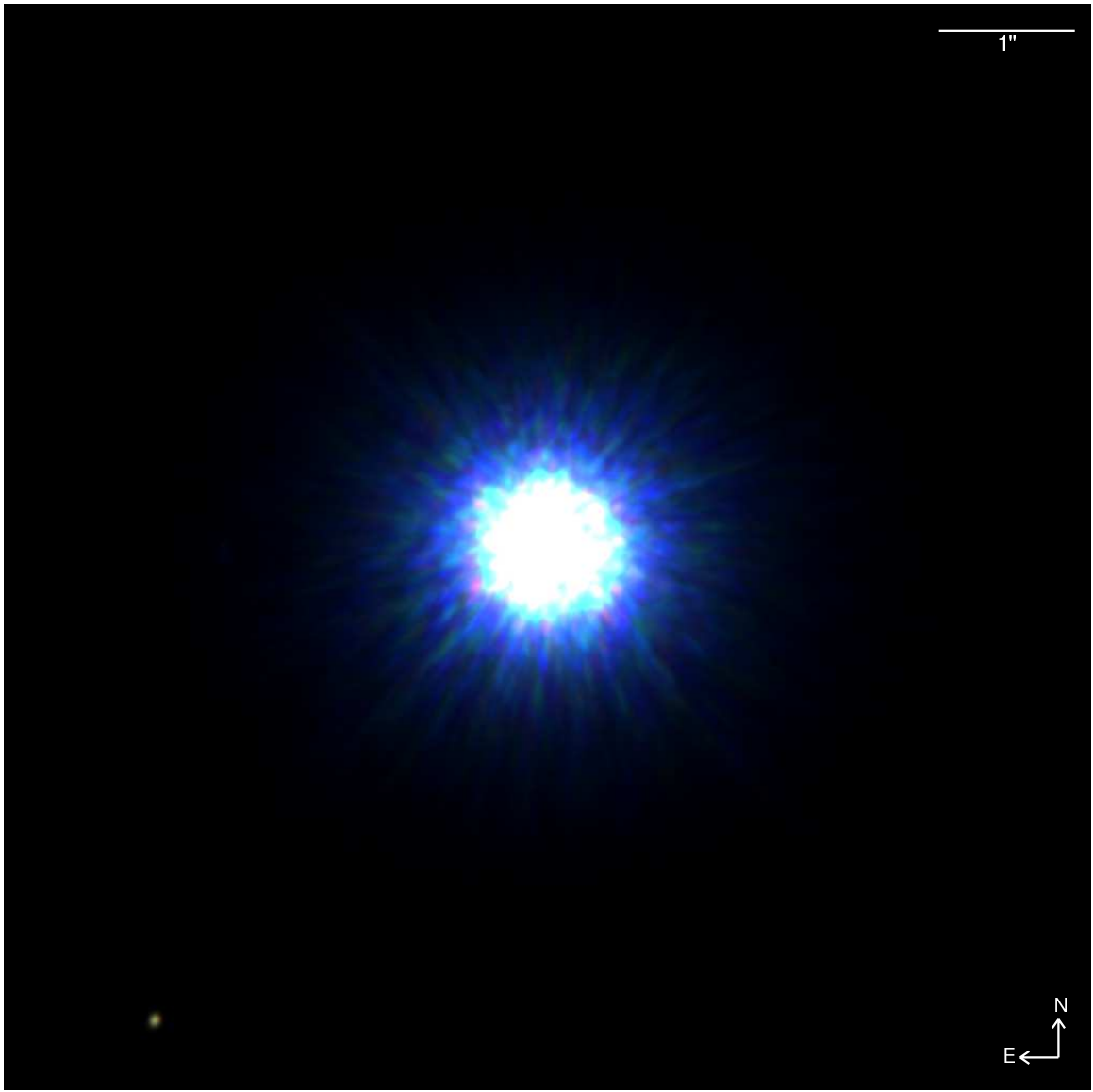}
\caption{\label{fig:im} Color composite $JHK$ image of \primary\ and its substellar companion (bottom left) obtained with NIRI/ALTAIR at the Gemini North telescope. The intensity scaling of each color is proportional to the incident photon flux in equal fractional bandwidths.}
\end{figure*}

The imaging data were reduced using custom {\em IDL} routines. For the 2008 and 2010 imaging data, a sky frame was constructed by taking the median of the images at all dither positions after masking out the regions dominated by the target's signal, while for the 2009 imaging data, the median of the three sky frames was obtained. After subtraction of this sky frame, the images were divided by a normalized flat-field. Then isolated bad pixels were replaced by the interpolated value of a third-order polynomial surface fit to the good pixels in a $7\times7$ pixels box while clustered bad pixels were simply masked out. Next the images were distortion corrected using the distortion solution provided by the Gemini staff.\footnote{The distortion correction used is centro-symmetric and is given by $r=r^\prime+1.32\times10^{-5}r^{\prime 2}$, where $r$ and $r^\prime$ are respectively the distortion-corrected and distorted radial pixel distances from the array center.} Finally, the long-exposure images were properly scaled in intensity and merged with their corresponding short-exposure images. The color composite $JHK$ image from the 2009 follow-up observations is shown in figure~\ref{fig:im}.

\subsection{Spectroscopy}

The likely companion detected in the imaging data was very faint and, assuming it was a true companion, its magnitude suggested a substellar mass. Thus, to verify its late-type nature, we obtained spectroscopic observations of it in $H$, $K$ and $J$ on  2009 July 2, 2009 July 3, and 2009 August 8, respectively, using the integral field spectrograph NIFS \citep{mcgregor03} with the ALTAIR adaptive optics system at the Gemini North telescope. The $J$ grating was used with the $ZJ$ blocking filter, the $H$ grating with the $JH$ blocking filter and the $K$ grating with the $HK$ blocking filter. The spectral resolving power is $\sim$6000 in $J$ and $\sim$5300 in both $H$ and $K$. Given the large angular separation of the companion ($\sim$4.5\arcsec), the primary star is not visible in the $3\arcsec\times 3\arcsec$ NIFS field-of-view. In each band, the companion was positioned near the center of the field-of-view and was dithered by 0.7\arcsec\ along a line between each of five exposures to enable good sky subtraction. The individual exposure times were 480~s, 300~s and 240~s in $J$, $H$ and $K$, respectively, in low read noise mode. After the $J$ and $H$ sequences, the A0V star HIP~79229 was observed at a similar airmass for telluric and instrumental transmission correction, while the A0V star HIP~73820 was observed just before the $K$ band sequence for the same purpose.
 
 \begin{deluxetable}{cccc}
\tablewidth{0pt}
\tablecolumns{4}
\tablecaption{Astrometric measurements \label{tbl:astrom}}
\tablehead{
\colhead{Epoch} & \colhead{Band} & \colhead{$\rho$ (\arcsec)} & \colhead{P.A. (deg)}
}
\startdata
2008.3940 & $K_{\rm cont}^{2.09}$ & $4.529\pm0.006$ & $140.32\pm0.10$ \\
2009.4998 & $JHK^\prime$              & $4.533\pm0.006$ & $140.35\pm0.10$ \\
2010.6605 & $K_{\rm cont}^{2.09}$ & $4.536\pm0.006$ & $140.27\pm0.10$
\enddata
\end{deluxetable}

The reduction of the NIFS data, up to the reconstruction of the data cube, was made using the Gemini IRAF pipeline. The steps covered in this pipeline are sky background subtraction, flat-field and bad pixel correction, spatial and spectral calibration, and data cube reconstruction. The spatial sampling of the reconstructed data cube is 0.043\arcsec\ pixel$^{-1}$. The instrumental/telluric transmission correction and spectrum extraction, which could be done using the Gemini pipeline, were instead done using custom IDL.

The center of each PSF was first registered to a common position in all spectral slices and all cubes of the sequence; the center positions were calculated by fitting a 2D Gaussian function. The spectrum of the source was extracted by summing the flux in a circular aperture of diameter 5 pixels. The spectrum of the telluric standard was corrected for its spectral slope using a 9520~K blackbody curve, and any hydrogen line absorption was removed by dividing out a Voigt profile fit. The companion spectrum was divided by the standard spectrum to correct for telluric and instrumental transmission. Finally, the median of the 5 spectra was obtained. 

\section{Analysis and results}\label{sect:analysis}

A summary of the properties of \primary A and its companion, taken from the literature or derived in this section, is presented in Table~\ref{tbl:properties}.

\subsection{Common proper motion}\label{sect:cpm}

\begin{figure}
\epsscale{1}
\plotone{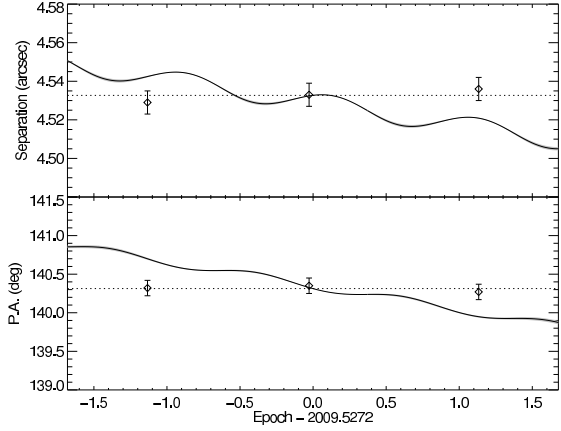}
\caption{\label{fig:pm}  Measured (data points) separations (top) and position angles (bottom) between the primary star and its companion as a function of time. The solid lines show the expected position of a distant stationary background source over time, as calculated from the proper motion and distance of the primary star.}
\end{figure}

The relative position of the companion and primary was determined by fitting a 2D Gaussian function to each component. The pixel positions were converted to arcseconds using the 21.4~mas pixel scale of NIRI with ALTAIR and the field lens as indicated on the instrument web page\footnote{\url{http://www.gemini.edu/sciops/instruments/altair/field-lens-option}}; the position angle of the image was taken from the FITS header. The astrometric errors, 6~mas in separation and 0.1\degr\ in position angle, were estimated in \citet{lafreniere10} using the position of background stars for a similar sequence of observations; they are mostly dominated by residual distortion errors. We also note that the pixel scale of ALTAIR/NIRI with the field lens has not been extensively calibrated and that our values may be systematically different from measurements made using other instruments or telescopes, but they should be internally consistent. The astrometry measurements are given in Table~\ref{tbl:astrom} and shown in Fig.~\ref{fig:pm} in comparison with the changes expected over time for a distant background star, based on the proper motion and distance of the primary star \citep[taken from][]{vanleeuwen07}. Our measurements are consistent with common proper motion but inconsistent, at a level of $\sim$6$\sigma$, with the changes expected for a distant, motionless background star. This clearly indicates that the companion is co-moving with the primary. 

The accuracy of our astrometry measurements can be verified using a faint background star detected both in 2008 and 2010. This source, $\sim$10.1~mag fainter than the primary, had a separation of 9.451\arcsec\ and position angle of 43.49\degr\ at epoch 2008.3940, and 9.493\arcsec\ and 43.48\degr, respectively, at epoch 2010.6605. These values, showing a $\sim$7$\sigma$ change in separation and $\sim$1$\sigma$ in position angle, are fully consistent with expectations for a distant, motionless background star.

The new companion we report in this paper was detected previously by \citet{kouwenhoven05} and \citet{kouwenhoven07} in data obtained in 2000 and 2001 with ADONIS at the ESO 3.6~m telescope. They report a separation of $4.54\arcsec\pm0.01\arcsec$ and position angle of $139.7\degr\pm0.3\degr$. They note that they were not able to conclude whether this object was a bound companion or a background object based on $HK$ photometry, and they have not followed up this source further. Considering the uncertainties involved in comparing our astrometry measurements with those made using other instruments, as noted above, the astrometry reported by Kouwenhoven~et~al. is in very good agreement with ours and further supports our conclusion that this source is co-moving with the primary star. Indeed, were the source a background star, its separation should have changed by $-0.1\arcsec$ and its position angle by $-3.3\degr$ between 2000 and 2010, while the measured values show changes of $+0.01\arcsec$ and $+0.6\degr$, respectively.

\subsection{Properties of the companion}

In this section, in addition to comparing our spectrum and photometry measurements with those of other objects, we also rely on comparisons with synthetic spectra to estimate the physical properties of the companion. For the comparison with models, we adopted the \driftph\ atmosphere models \citep{witte09, helling08, dehn07}, which combine a detailed kinetic model of dust cloud formation with a radiative transfer code \citep{hauschildt99, baron03}. In contrast to all other atmosphere models which assume phase equilibrium between gas and cloud particles, \driftph\ describes consistently the formation of a stationary cloud by homogeneous nucleation and grain growth/evaporation, including gravitational settling, element depletion, and convective element replenishment \citep{woitke03, woitke04, helling06}. All models used assume solar metallicity, which is reasonable for Upper~Sco \citep{mohanty04}.

\begin{deluxetable}{lcc}
\tablewidth{0pt}
\tablecolumns{3}
\tablecaption{\label{tbl:properties} Properties of \primary\ AB}
\tablehead{
\colhead{} & \multicolumn{2}{c}{Value} \\
\cline{2-3}
\colhead{Parameter} & \colhead{Primary} & \colhead{Companion}
}
\startdata
$\mu_\alpha \cos{\delta}$ (mas yr$^{-1}$)\tablenotemark{a}& $-11.38\pm0.59$ & \nodata \\
$\mu_\delta$ (mas yr$^{-1}$)\tablenotemark{a}& $-24.66\pm0.37$ & \nodata \\
Distance (pc)\tablenotemark{a} & $156.7\pm13.0$ & \nodata \\
Angular separation (\arcsec) & \multicolumn{2}{c}{$4.533\pm0.006$} \\
Position angle (deg) & \multicolumn{2}{c}{$140.3\pm0.1$} \\
$\Delta J$ (mag)         & \multicolumn{2}{c}{$8.13\pm0.05$} \\
$\Delta H$ (mag)         & \multicolumn{2}{c}{$7.44\pm0.03$} \\
$\Delta K^\prime$ (mag) & \multicolumn{2}{c}{$7.27\pm0.03$} \\
$\Delta K^{\rm 2.09}_{\rm cont}$ (mag) & \multicolumn{2}{c}{$7.26\pm0.03$} \\
$J$ (mag)\tablenotemark{b}        & $6.928\pm0.021$ & $15.06\pm0.05$ \\
$H$ (mag)\tablenotemark{b}        & $6.946\pm0.029$ & $14.39\pm0.04$ \\
$K_{\rm s}$ (mag)\tablenotemark{b} & $6.903\pm0.020$ & $14.17\pm0.04$ \\
$J-K_{\rm s}$ (mag) & $0.025\pm0.03$ & $0.89\pm0.06$ \\
$H-K_{\rm s}$ (mag) & $0.043\pm0.04$ & $0.22\pm0.06$ \\
Spectral type & B9V & M8$\pm1$ \\
$T_{\rm eff}$ (K) & $\sim$10500\tablenotemark{c} & 2800$\pm200$ \\
$\log{(L/L_\odot)}$ & \nodata & $-2.55\pm0.13$ \\
Mass ($M_\odot$) & $\sim$2.5\tablenotemark{d} & $0.022\pm0.004$\tablenotemark{e} \\
Projected separation (AU) & \multicolumn{2}{c}{$710\pm60$}
\enddata
\tablenotetext{a}{From \citet{vanleeuwen07}.}
\tablenotetext{b}{From the 2MASS PSC and our contrast measurements. Differences in filter bandpasses may add up to 1-2\% uncertainty.}
\tablenotetext{c}{From the spectral type based on the temperature scale of \citet{sherry04}.}
\tablenotetext{d}{From the models of \citet{dantona97}.}
\tablenotetext{e}{From the models of \citet{baraffe98, baraffe02} and \citet{burrows97}.}

\end{deluxetable}

\subsubsection{Temperature, surface gravity and spectral type}

\begin{figure*}
\epsscale{1}
\plotone{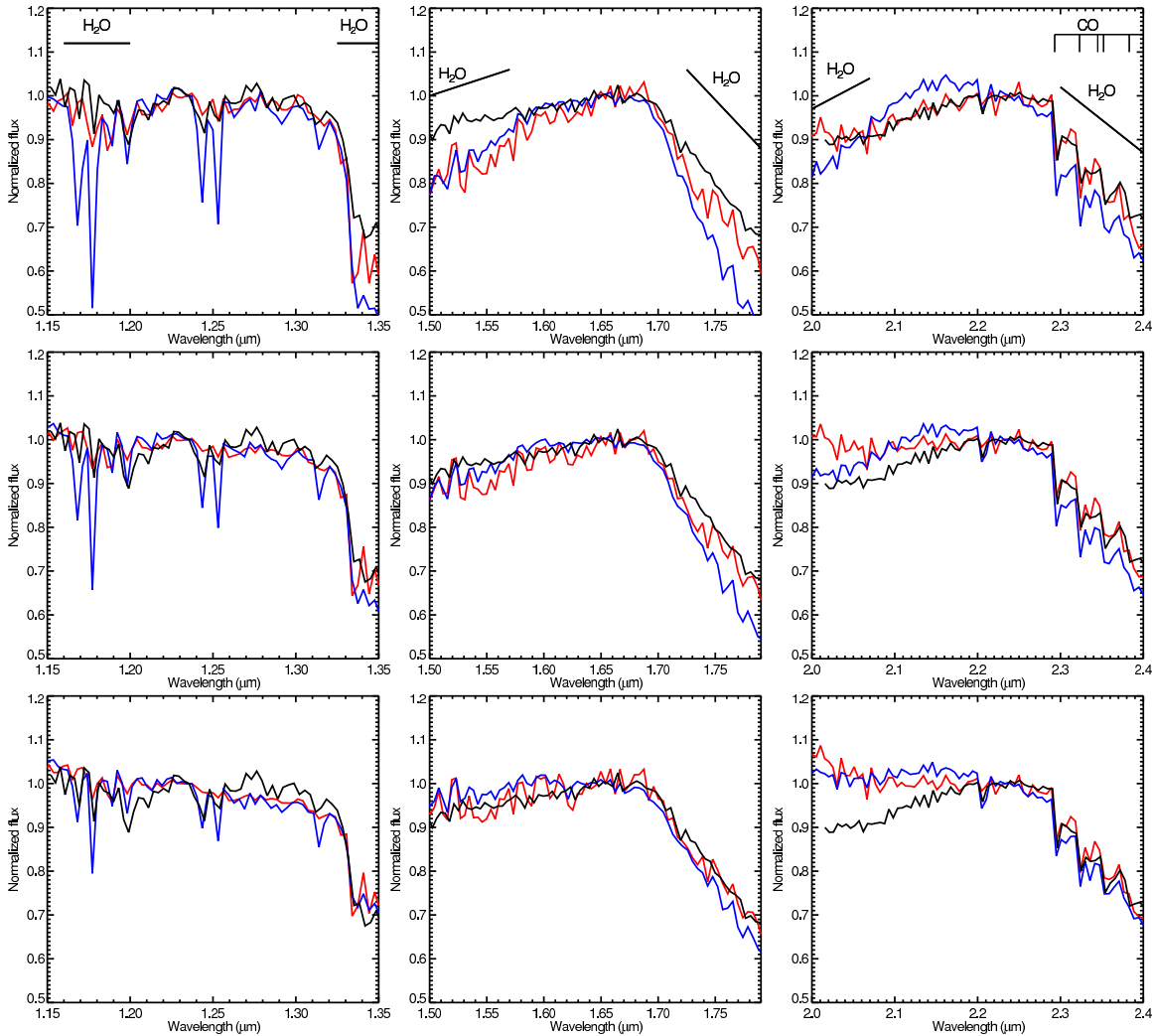}
\caption{\label{fig:spec1} NIFS spectrum of \comp\ (black) compared with synthetic spectra from the \driftph\ atmosphere models (red and blue) of various effective temperatures and surface gravities. From the top to the bottom rows, the effective temperatures are 2600, 2800 and 3000~K, respectively. The red curves are for $\log{g}=4.0$ and the blue curves are for $\log{g}=6.0$. The spectra were binned to a resolving power of 400 and were normalized separately in each spectral band.}
\end{figure*}

The spectrum of \comp, binned down to a resolving power of 400, is shown in figure~\ref{fig:spec1} in comparison with synthetic spectra from the \driftph\ atmosphere models for various temperatures and surface gravities. As visible on the figure, for all effective temperatures shown, the spectrum of the companion is in better agreement with the lower surface gravity spectrum, as expected for a young object in Upper~Sco. This is particularly striking in the $K$ band where the broad H$_2$O absorption bands and the CO absorption band head depths are perfectly reproduced by the low gravity spectrum at $T_{\rm eff}=2600$~K. The low surface gravity is also apparent in the depth of the K{\sc i} lines in the $J$ band. The effective temperature providing the best fit depends on the bandpass considered. As mentioned previously, a temperature of 2600~K provides an excellent fit in the K band. A temperature of 2600~K also provides a slightly better fit in the $J$ band, both for the continuum around 1.2~$\mu$m and the depth of the K{\sc i} lines. In the $H$ band, however, the best fit is obtained for temperatures of $\sim$2800-3000~K. Taking these considerations into account, we adopt an effective temperature of $2800\pm200$~K. We have also done a chi-square minimization using the same set of model spectra, rather than a simple visual inspection, and reached the same conclusion. In particular, when the minimization is done in all bands simultaneously, the best fit is found for a temperature of 2800~K.

\begin{figure*}
\epsscale{1}
\plotone{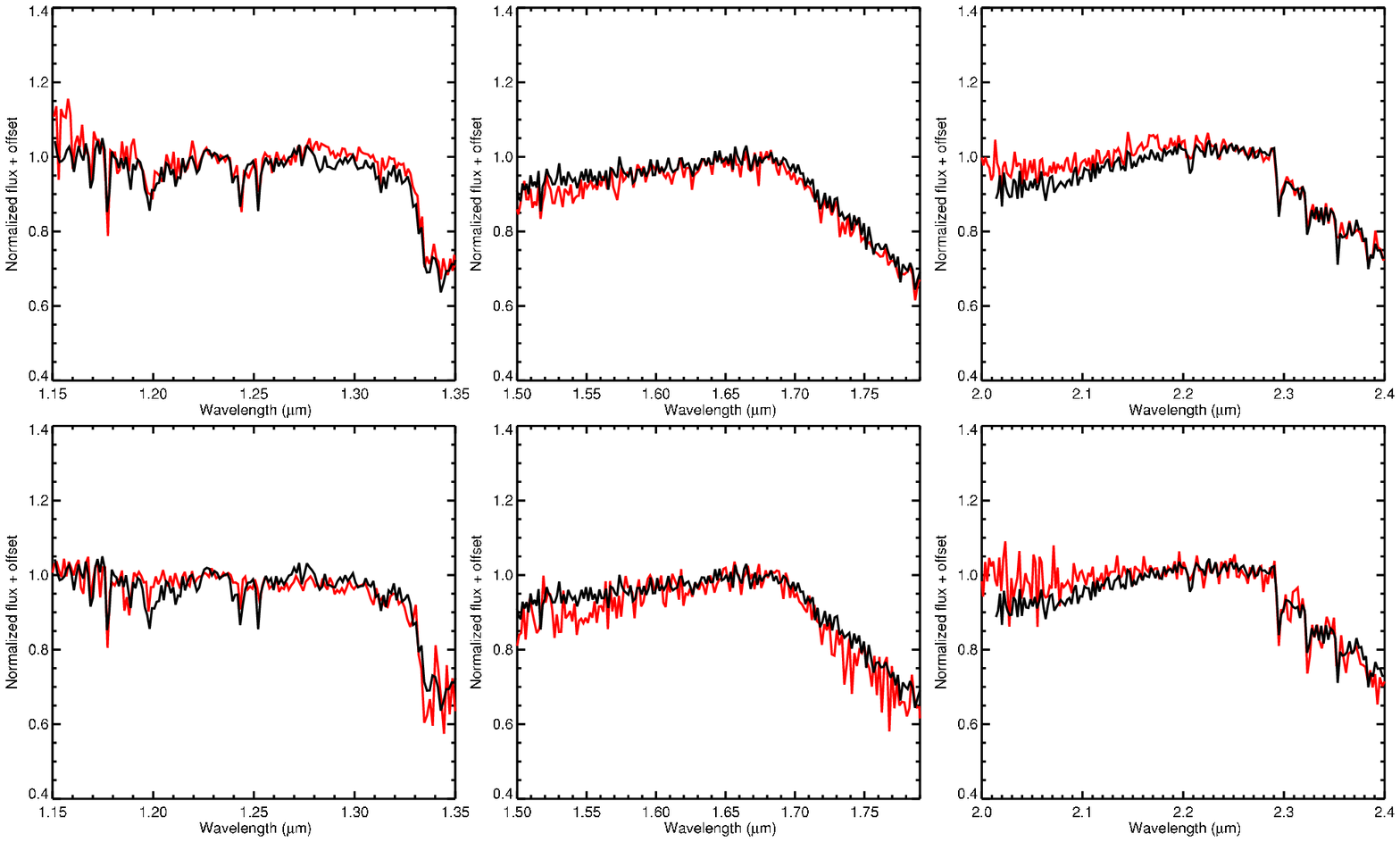}
\caption{\label{fig:spec2} NIFS spectrum of \comp\ (black) compared with the spectrum of USco~J155419.99-213543.1 (top), which is an M8 free-floating brown dwarf member of Upper~Sco \citep{lodieu08}, and with a synthetic spectrum from the \driftph\ atmosphere models with $\teff=2800$~K, $\log{g}=4.5$, and solar metallicity (bottom). The spectra were binned to a resolving power of 1000 and were normalized separately in each spectral band.}
\end{figure*}

Figure~\ref{fig:spec2} shows the companion spectrum, at a higher resolution this time, in comparison with the spectrum of a known M8 dwarf member of Upper Sco as well as with a model spectrum with our effective temperature estimate of 2800~K. Among the Upper~Sco brown dwarfs identified and spectrally classified by \citet{lodieu08}, the best fit to our spectrum is obtained for a spectral type of M8, which corresponds to an effective temperature of $\sim$2700-2800~K, consistent with our previous estimate. Comparing our spectrum to those of field dwarfs from the IRTF spectral library \citep{rayner09} also results in a best fit for spectral types of M7-M9, although there are differences between the spectra owing to the lower gravity of the companion as noted earlier. The better agreement of our spectrum with those of other objects in Upper Sco, as opposed to field objects, provides further evidence that the new companion also belongs to the association.

The companion spectrum is not only in good agreement with those of models and other Upper Sco objects within the individual bands, but it also shows a reasonable agreement in colors between the bands. Over a temperature range of 2600--3000~K, the \driftph\ models yield synthetic colors of $J-H$=0.50-0.55 and $H-K$=0.27-0.38, to be compared with the companion colors of $J-H$=$0.67\pm0.06$ and $H-K$=$0.22\pm0.06$. Assuming a small possible extinction of $A_{\rm V}=0.5$~mag toward this system \citep{carpenter09}, the extinction-corrected colors would be $J-H$$\sim$0.62 and $H-K$$\sim$0.19. The colors of M8 dwarfs in Upper Sco from the sample of \citet{lodieu08} are 0.58-0.65 and 0.38-0.57, respectively. Here, the agreement in $J-H$ is good but our companion is significantly bluer in $H-K$. We note that the differences between the NIRI (used for the companion contrast) and 2MASS (used for the primary star magnitude) filter bandpasses add an uncertainty of at most 2-3\% on the colors of the companions, calculated from synthetic spectra.

\begin{figure*}
\epsscale{1}
\plotone{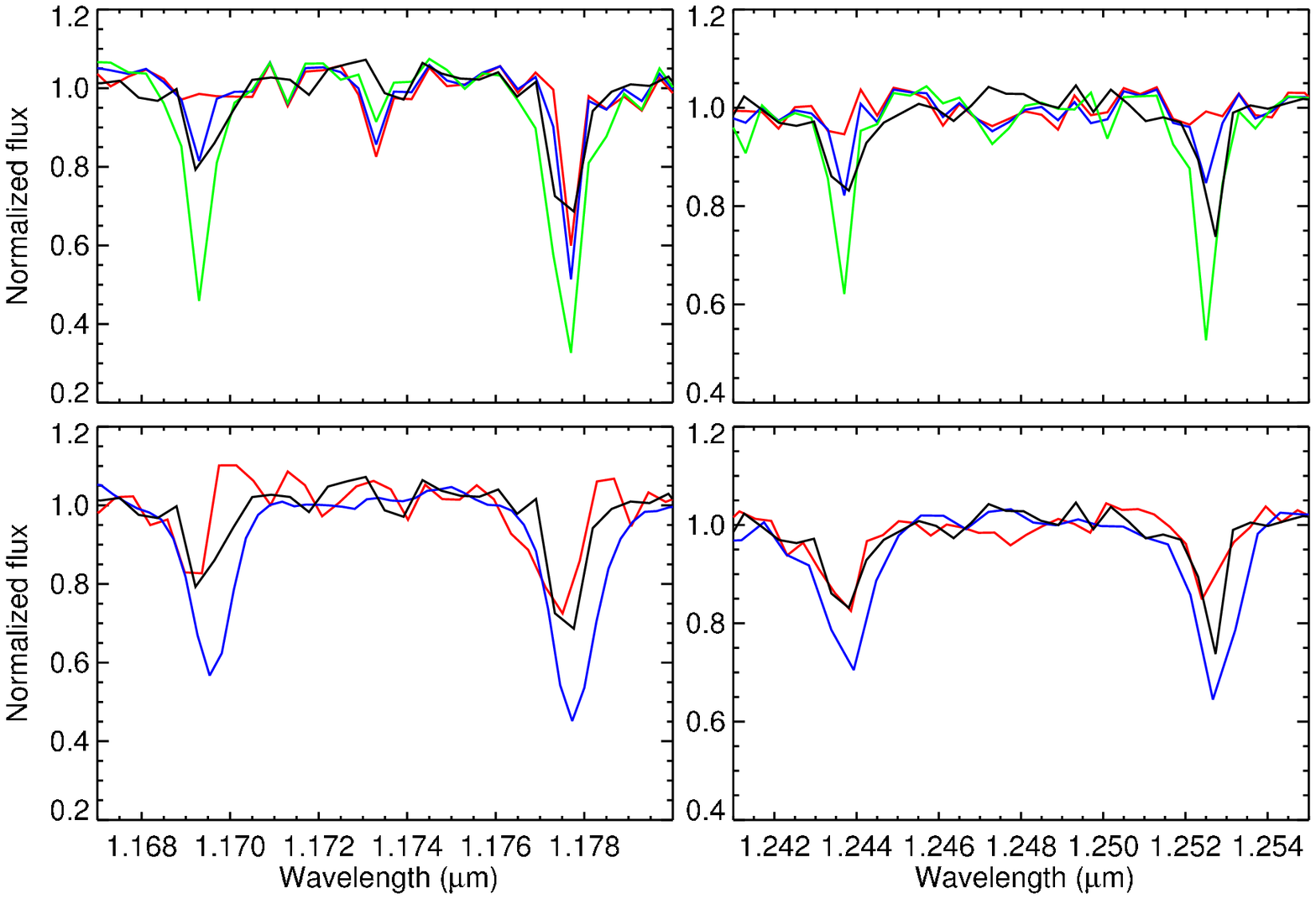}
\caption{\label{fig:KI} Spectrum of \comp\ (black) showing the K{\sc i} doublets at $\sim$1.17~$\mu$m and $\sim$1.25~$\mu$m. In the top panels, the companion spectrum is compared with spectra from the \driftph\ models for $T_{\rm eff}=2800$~K and $\log{g}=3.5$(red), 4.5 (blue), and 5.5 (green). In the bottom panels, the spectrum is compared with the spectrum of USco~J155419.99-213543.1 (red, \citealt{lodieu08}), a young M8 brown dwarf in Upper~Sco, and LP412-31 (blue, \citealt{mclean03}), a field M8 dwarf. The shallower K{\sc i} lines seen in the companion spectrum relative to the field object indicates lower surface gravity. The spectra were binned to a resolving power of $\sim$3000.}
\end{figure*}

A good indicator of surface gravity is the equivalent width (EW) of the K{\sc i} atomic line doublets at 1.168/1.177~$\mu$m and 1.243/1.253~$\mu$m \citep[e.g.][]{mcgovern04}.
For \comp, the EWs of these four lines are, respectively, $2.0\pm0.3$~\AA, $3.0\pm0.3$~\AA, $2.0\pm0.2$~\AA, and $2.1\pm0.2$~\AA.\footnote{To calculate the EWs, we used an integration interval of 10~nm centered on each line and fitted the continuum as a straight line using 40~nm-wide intervals on both sides of the line.}  In the \driftph\ models, there is a clear monotonic trend that the EW in these lines increases with increasing surface gravity for a constant temperature, see figure \ref{fig:KI}. We have evaluated the EW for models of 2800~K with surface gravity ranging from 3.5 to 6.0, and interpolated the measured EW of the companion for each respective line. The mean and standard deviation of the four lines from this comparison indicate a surface gravity of $\log{g} = 4.6 \pm 0.3$, which is relatively low, as expected and verified for young objects\citep[e.g.][]{brandeker06}. The K{\sc i} EWs we measured for the companion are also comparable to those of other M8-M9 members of Upper Sco (using spectra from \citealp{lodieu08}) but only a third to half of the EWs of M7-M9 field dwarfs of comparable spectral types (using spectra from \citealp{mclean03}), clearly indicating low surface gravity and membership in Upper Sco for the new companion (see also figure~\ref{fig:KI}).

\subsubsection{Atmosphere composition and structure}

\driftph\ provides local gas properties like the gas temperature T [K], the gas density $\rho$ [g\, cm$^{-3}$], and the local gas-phase composition, but also dust quantities such as the number of dust particles $n_{\rm d}$ [cm$^{-3}$] of mean grain size  $\langle a \rangle$ [cm] at each layer of the atmosphere. These quantities are needed to evaluate the radiation transfer through the atmosphere taking into account convective energy transport, hence to calculate the synthetic spectrum as shown in Fig.~\ref{fig:spec1}.

\begin{figure}
\epsscale{1}
\plotone{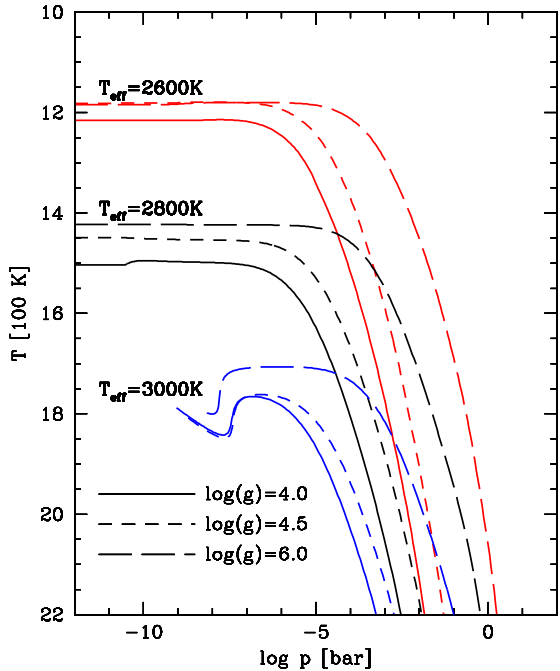}
\caption{\label{fig:tp} The temperature--pressure structures for all \driftph\ atmosphere models used in Fig.~\ref{fig:spec1}. The $T-P$ profiles vary considerably for the three sets of models at different $T_{\rm eff}$, although the emerging spectra from the three low-gravity atmosphere all provide a reasonable agreement with the observed companion spectrum.}
\end{figure}

The local gas-phase and cloud chemistry is determined by the local temperature and pressure which we demonstrate in Fig.~\ref{fig:tp} for all \driftph\ atmosphere models applied in Fig.~\ref{fig:spec1}. While the low gravity spectra in the 2600-3000~K temperature range all provide a reasonable agreement with the observed spectrum, figure~\ref{fig:tp} shows that the corresponding underlying atmospheric $(T, p)$ profiles are very different. This emphasises that medium differences in the spectral energy distribution can hide large variation of the atmosphere's structure. The models at 3000~K even develop a small temperature inversion high up in the atmosphere, where H$_2$O is dissociated, freeing up oxygen and locally enhancing the opacity compared to the surrounding layers.\footnote{Such a temperature and pressure inversion does not necesserely invalidate the assumption of hydrostatic equilibrium which is applied in the \driftph\ atmosphere models as was shown in \citet{helling00} and in \citet{asplund98}.} 

\begin{figure}
\epsscale{1}
\plotone{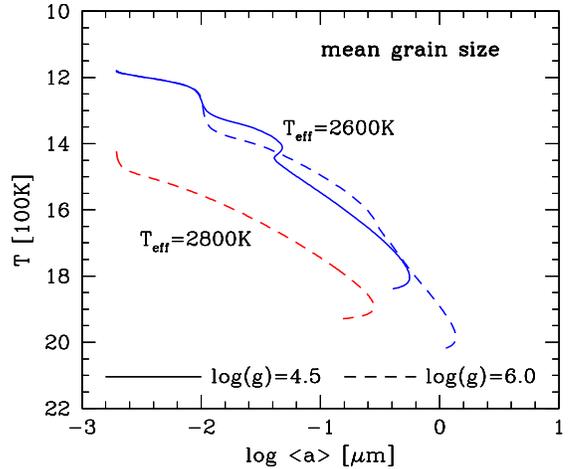}
\caption{\label{fig:dust} The mean grain sizes across the clouds altitudes in the three dust-forming model atmospheres plotted in Fig.~\ref{fig:spec1}. The cloud grains remain rather small in all models and reside in the optical thin region of the atmosphere.}
\end{figure}

Of all the models shown on figure~\ref{fig:spec1}, only three do have dust in their atmosphere: $\log{g}=4.5$ with $T_{\rm eff}=2600$~K and $\log{g}=6.0$ with both $T_{\rm eff}=2600$~K and 2800~K. Figure~\ref{fig:dust} shows the mean grain sizes for these models as function of local temperature; the lowest temperatures indicate the upper part of the atmosphere. The dust persists at a higher temperature in the high-gravity models owing to an increased local density; however, these models are less relevant for the new companion found here, which is young and has low gravity. Thus unless the companion is on the low end of our estimated $T_{\rm eff}$ range, it probably does not have dust in its atmosphere. However if it is indeed closer to 2600~K and has dust in its atmosphere, the grain sizes would remain rather small,  forming a haze layer in the upper, optically thin layers of the atmospheres (see Fig.~\ref{fig:dust}). The only objects where haze layers have been inferred from transit photometry, however, are irradiated giant gas planets (see \citet{sing09}). No direct detection of haze on brown dwarfs has been obtained to date.

\subsubsection{Luminosity and mass}

We estimated the bolometric luminosity of the companion by using its observed photometry in combination with model spectra. We first computed synthetic fluxes in the $JHK$ bands using the model spectra, then adjusted those synthetic fluxes to the measured values, and finally integrated the scaled model spectra over all wavelengths to obtain the total irradiance, which was then converted to bolometric luminosity using the primary star distance. The synthetic average flux densities in the $JHK$ bands were computed using the relative spectral response curves given on the 2MASS project webpage\footnote{\url{http://www.ipac.caltech.edu/2mass/releases/allsky/doc/sec6\_4a.html}}. We repeated the procedure using both the \driftph\ and NextGen \citep{hauschildt99} model spectra for ranges of $T_{\rm eff}$ from 2600~K to 3000~K and $\log{g}$ from 3.5 to 5.0. We obtained a value of $\log{(L/L_\odot)}=-2.55\pm0.13$. The uncertainty quoted reflects the range of values obtained for the different model parameters and accounts for the uncertainty on the distance of the primary star.

\begin{figure}
\epsscale{1}
\plotone{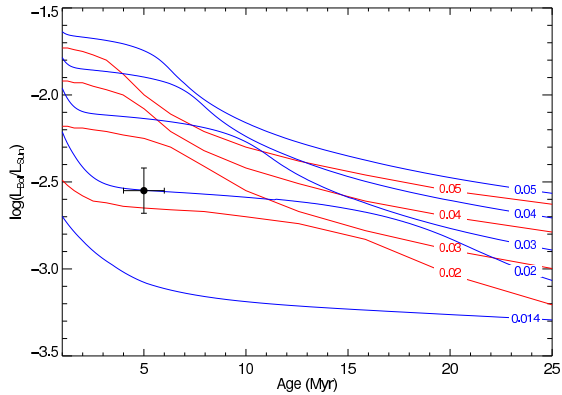}
\caption{\label{fig:lbol} Calculated bolometric luminosity of the companion (black point) compared with evolutionary tracks from \citet{baraffe98, baraffe02} (red curves) and \citet{burrows97} (blue curves). The curves are labeled with their mass expressed in units of solar mass.}
\end{figure}

The age of the Upper Scorpius association is well constrained at 5~Myr and all stars appear to have formed in a burst, over a period of at most 1-2~Myr \citep{preibisch02,slesnick08}. For an age of $5\pm1$~Myr and the above companion bolometric luminosity estimate, the evolution models of \citet{chabrier00} indicate a mass of $\sim$21-26~\mjup, while those of \citet{burrows97} yield a mass of $\sim$19-25~\mjup, see figure~\ref{fig:lbol}. The ranges of mass quoted reflect only the uncertainties on the age and luminosity estimates, but of course, larger uncertainty may be present owing to the lack of absolute calibration of the evolution models. As visible on figure~\ref{fig:lbol}, the companion is currently in a phase where its luminosity is more or less constant with time due to deuterium burning; this phase will last until an age of $\sim$15~Myr. Interestingly, this means that the companion mass estimate is not strongly dependent on the age estimate, as is ordinarily the case for substellar objects.

\section{Discussion and concluding remarks}\label{sect:discussion}

Due to the very large semi-major axis of the companion orbit, its expected orbital motion per year is below the astrometric precision of our data. To estimate the orbital motion, we can consider the hypothetical case in which the orbit is circular and face-on as seen from Earth, with a semi-major axis of 710~AU. Given the system mass of approximately 2.5~\msun, this yields an orbital period of $\sim$12000 years, which in turn gives an angular motion of 0.03\degr~yr$^{-1}$ (or a linear motion of $\sim$2.4~mas~yr$^{-1}$). The angular motion over a decade under this assumption, 0.3\degr, does come out to the same order of magnitude as the 2$\sigma$ difference in position angle between our images and those of \citet{kouwenhoven05}, $0.6\degr\pm0.3$\degr. However, in reality, the orbital motion is likely to be much smaller, as the projected separation is typically smaller than the real semi-major axis. In addition, the only way in which the angular motion could be larger than 0.3\degr\ is if the orbit is eccentric and the companion is presently close to periastron, which is a priori unlikely. As we note in \S\ref{sect:cpm}, our astrometry is internally consistent for the NIRI data but not necessarily with respect to other instruments, hence it is plausible that the difference with Kouwenhoven et al. is simply spurious. Thus, there is no convincing evidence of orbital motion, and no such motion is expected over the relevant timescales.

\begin{figure}
\epsscale{1}
\plotone{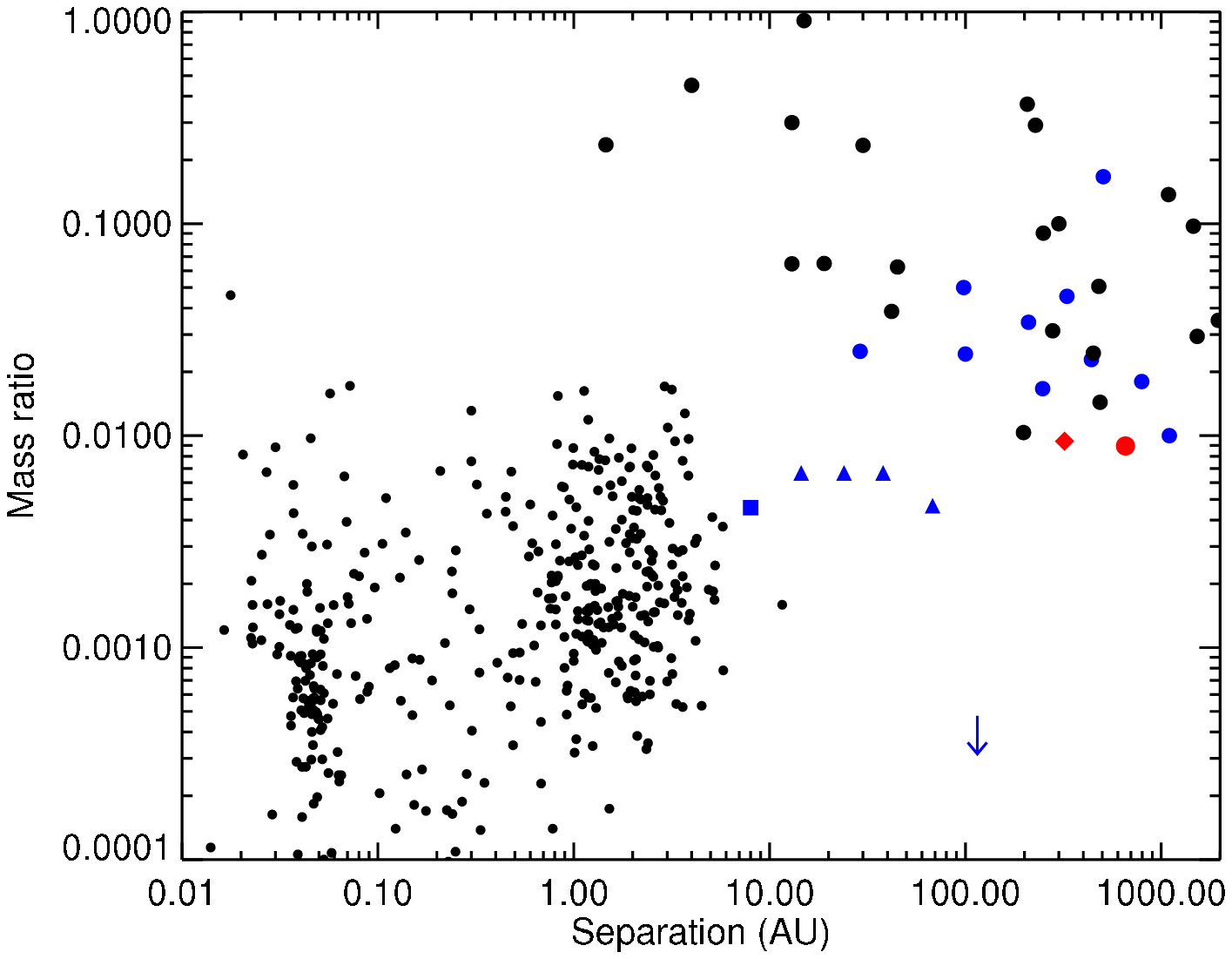}
\caption{\label{fig:qvssep} Mass ratio as a function of separation for various substellar companions to stars. The red filled circle is \comp\ and the red filled diamond is 1RXS~J160929.1-210524b. The blue filled circles are all other directly imaged low-mass substellar companions ($<25$~\mjup) to stars from the compilation given in \citet{lafreniere10}, augmented with the newly-found companion Ross~458C \citep{goldman10, scholz10}. The large black filled circles are more massive directly imaged substellar companions from the compilation of \citet{zuckerman09}. The triangles, square and down arrow are HR~8799bcde \citep{marois08, marois10}, $\beta$~Pic b \citep{lagrange09a, lagrange10} and Fomalhaut~b \citep{kalas08}, respectively. The small black circles indicate planets found by the radial velocity and microlensing techniques from the Extrasolar Planets Encyclopaedia (http://exoplanet.eu/).}
\end{figure}

A mass ratio--separation diagram showing \comp\ and other known brown dwarf and planetary companions to stars is shown in Figure~\ref{fig:qvssep}. With a mass of $\sim$23~\mjup\ and projected separation of $\sim$700~AU from its massive 2.5~\msun\ primary, \comp\ lies at the lower boundary of all known companions with separations larger than 100~AU. It is unclear whether these companions, whose mass ratios overlap with those of both more massive brown dwarf companions and bona fide planets, formed in a planet-like or in a stellar-like manner. Given the size of their orbits, it appears unlikely that they formed in situ in a planet-like manner -- either through the collapse of a gravitationally unstable disk \citep[e.g.][]{vorobyov10} or by the core accretion mechanism \citep[e.g.][]{pollack96}. Indeed for both scenarios the disks would need to be unusually large and massive, but even then, the formation timescale by core accretion would be prohibitively long while the efficiency of disk instability to produce such companions is highly uncertain \citep{vorobyov10, kratter10}. Nevertheless, it is possible that these companions did form in a planet-like manner much closer to the star, but were subsequently kicked outward through gravitational interactions \citep[e.g.][]{scharf09,veras09}. In this scenario, according to the simulations of \citet{veras09}, the timescale for instabilities to develop and send planets on large orbits may be quite short (0.01-1~Myr), with significant dynamical evolution occurring within the first few Myr. However, owing to further evolution, planets quickly scattered on large orbits are likely to be ejected from the system after a few tens of Myr. Alternatively, these wide and low mass ratio companions could form like stellar binaries, through the fragmentation of a pre-stellar core \citep[e.g][]{bate09,bate03}. Numerical simulations indicate that very low mass ratio companions can indeed be produced by this process, albeit rarely, and that they preferentially have large separations and are generally found in high order multiple systems. Future observations may provide constraints to exclude or support the various formation possibilities of these wide low mass companions. For instance, if they formed in a planet-like manner closer to the star and were subsequently kicked outward through gravitational interactions, then it would be expected that additional companions of similar mass or even heavier should be present in the systems, at smaller separations.

In addition, as mentioned above, distant companions produced through this process would likely be eventually ejected from the system, such that they should be found more frequently in star-forming regions than around older stars. In our overall survey of Upper Scorpius that led to the discovery of \comp\ and 1RXS~J1609-2105b, we observed 91 stars (masses 0.15-5~\msun), so taken at face value, our two discoveries imply that companions with mass ratios below 0.01 and separations of hundreds of AUs exist in 2.2$_{-1.9}^{+5.5}$\% (95\% credibility) of stellar systems. Considering that we did not achieve the same sensitivity for all targets as well as our incompleteness to lower mass ratio companions, this number is only a lower limit. The statistics are not yet sufficient to tell whether there exists a difference for older objects. For example the study of \citet{lafreniere07}, which targeted $\sim$200~Myr-old GKM stars, enabled placing upper limits of $\sim$6\% (95\% credibility) for companions of $\sim$10~\mjup\ (i.e. mass ratio $\sim$0.01); see also \citet{chauvin10} and \citet{nielsen10}. Improving the statistics for both young and older systems would thus be a good way to investigate further the importance of gravitational scattering in accounting for distant companions. This approach is valid as long as the internal dynamics of the multiple planets dominates over interactions with other stars or giant molecular clouds as the system travels through the galaxy, as these interactions can also strip out wide companions. This can be roughly checked using the results of \citet{weinberg87}. Based on their figure~2, with $a/M_{\rm tot}\sim0.0014$~pc~\msun$^{-1}$, the disruption lifetime of HIP~78530AB due to encounters with stars and giant molecular clouds is larger than 10~Gyr; this is also the case for 1RXS~J160929.1--210524Ab. In other words, stellar and molecular clouds encounters have little impact on the evolution of companions such as the ones we have found.

The new companion reported in this paper joins a growing list of low-mass substellar companions ($\la$25~\mjup) in wide orbit ($\ga$100~AU) around stars, which now counts about ten objects (see \citet{lafreniere10} for a recent compilation, see also Fig.~\ref{fig:qvssep}). These companions are found around primaries covering a wide range of masses and even around primaries that are themselves binaries. Thus, while the statistics on their frequency may not be very accurate, it seems likely that wide low-mass substellar companions are not an unusual outcome of the star formation process, yet they remain hard to explain within current theoretical frameworks.  Over the next few years, additional searches for new companions and continued efforts to characterize the known ones should allow us to make good progress toward explaining their formation.

\acknowledgments

We thank the Gemini staff for help and support with the observations. 
The authors also wish to thank Marten van Kerkwijk, Alexis Brandeker, Christian Marois and \'Etienne Artigau for useful discussion or help regarding some aspects of this work.
RJ acknowledges support from NSERC grants and a Royal Netherlands Academy of Arts and Sciences (KNAW) visiting professorship.
Finally, we thank our referee, Dr. Kevin Luhman, for helpful suggestions to improve this paper.


\end{document}